\documentstyle[11pt,newpasp,twoside,epsf]{article}
\def\edcomment#1{\iffalse\marginpar{\raggedright\sl#1\/}\else\relax\fi}
\reversemarginpar

\newcommand{\vy}[2]{#1_{\scriptscriptstyle #2}}
\newcommand{\Ly}{Ly$\alpha$}

\def\gtorder{\mathrel{\raise.3ex\hbox{$>$}\mkern-14mu
             \lower0.6ex\hbox{$\sim$}}}
\def\ltorder{\mathrel{\raise.3ex\hbox{$<$}\mkern-14mu
             \lower0.6ex\hbox{$\sim$}}}
\def\proptwid{\mathrel{\raise.3ex\hbox{$\propto$}\mkern-14mu
             \lower0.6ex\hbox{$\sim$}}}
\def\0946{PG~0946+301}

\def\arcsec{\ifmmode '' \else $''$\fi}

\def\arcsecpoint{\ifmmode ''\!. \else $''\!.$\fi}

\def\kms{\ifmmode {\rm km\ s}^{-1} \else km s$^{-1}$\fi}
\def\Msun{\ifmmode {\rm M}_{\odot} \else M$_{\odot}$\fi}
\def\Lsun{\ifmmode {\rm L}_{\odot} \else L$_{\odot}$\fi}
\def\Zsun{\ifmmode {\rm Z}_{\odot} \else Z$_{\odot}$\fi}

\def\ergscm2{ergs\,s$^{-1}$\,cm$^{-2}$}
\def\icm3{{\rm cm}^{-3}}
\def\icm2{{\rm cm}^{-2}}
\def\qo{\ifmmode q_{\rm o} \else $q_{\rm o}$\fi}
\def\Ho{\ifmmode H_{\rm o} \else $H_{\rm o}$\fi}
\def\ho{\ifmmode h_{\rm o} \else $h_{\rm o}$\fi}

\def\vFWHM{\ifmmode v_{\mbox{\tiny FWHM}} \else
            $v_{\mbox{\tiny FWHM}}$\fi}
\def\CCF{\ifmmode F_{\it CCF} \else $F_{\it CCF}$\fi}
\def\ACF{\ifmmode F_{\it ACF} \else $F_{\it ACF}$\fi}
\def\Halpha{\ifmmode {\rm H}\alpha \else H$\alpha$\fi}
\def\Hbeta{\ifmmode {\rm H}\beta \else H$\beta$\fi}
\def\Hgamma{\ifmmode {\rm H}\gamma \else H$\gamma$\fi}
\def\Hdelta{\ifmmode {\rm H}\delta \else H$\delta$\fi}
\def\Lya{\ifmmode {\rm Ly}\alpha \else Ly$\alpha$\fi}
\def\Lyb{\ifmmode {\rm Ly}\beta \else Ly$\beta$\fi}
\def\Lyg{\ifmmode {\rm Ly}\beta \else Ly$\gamma$\fi}

\def\ciii{\ifmmode {\rm C}\,{\sc iii} \else C\,{\sc iii}\fi}
\def\civ{\ifmmode {\rm C}\,{\sc iv} \else C\,{\sc iv}\fi}

\def\nv{N\,{\sc v}}

\def\o5007{[O\,{\sc iii}]\,$\lambda5007$}

\def\ovi{O\,{\sc vi}}

\def\o{\o}
\begin{document}
\title{OUTFLOWS VS. CLOUDS IN AGN \\
INTRINSIC ABSORBERS}
\author{Nahum Arav}
\affil{CASA, 
University of Colorado 
389 UCB 
Boulder, CO 80309-0389 
}

\begin{abstract}

We discuss the crucial role of a dynamical picture in the analysis of
AGN intrinsic absorbers data.  High quality FUSE data of Mrk 279 are
used to demonstrate that the line of sight covering fraction is a
strong function of velocity.  In Mrk 279, as well as in most cases
where the data is of high enough quality, the shape of the absorption
troughs is mainly determined by the velocity-dependent covering
fraction. We argue that the traditional ``cloud'' picture of AGN
outflows is hard pressed to account for the velocity-dependent
covering fraction, as well as for the highly super-thermal width of the
troughs and the detached trough phenomenon.  A disk outflow picture
naturally explains these features and furthermore, is using the
simplest reservoir for the outflowing material: The accretion disk
around the black hole. Accounting for velocity dependent covering
can drastically increase the inferred ionic column density of the
analyzed trough, an increase which is amplified in the total column
density and ionization parameter solution for the AGN outflow trough.

\end{abstract}

\section{Introduction}
AGN outflows are evident by resonance line absorption troughs, which
are blueshifted with respect to the systemic redshift of their
emission counterparts. In Seyfert galaxies, velocities of several
hundred \kms\ (Crenshaw et~al.\ 1999; Kriss et~al.\ 2000) are observed
in both UV resonance lines (e.g.,
\civ~$\lambda\lambda$1548.20,1550.77,
\nv~$\lambda\lambda$1238.82,1242.80,
\ovi~$\lambda\lambda$1031.93,1037.62 and \Ly), as well as in X-ray
resonance lines (Kaastra et~al.\ 2000; Kaspi et~al.\ 2000). Similar
outflows (often with significantly higher velocities) are seen in
quasars (Weymann et~al.\ 1991; Korista, Voit, Morris, \& Weymann 1993;
Arav et~al.\ 2001a). Reliable measurement of the absorption column
densities in the troughs are crucial for determining the ionization
equilibrium and abundances of the outflows, and the relationship
between the UV and the ionized X-ray absorbers.

In the last few years our group (Arav 1997; Arav et~al.\ 1999a; Arav
et~al.\ 1999b; de~Kool et~al.\ 2001; Arav et~al.\ 2001a) and others
(Barlow 1997, Telfer et~al.\ 1998, Churchill et~al.\ 1999, Ganguly
et~al.\ 1999) have shown that in quasar outflows most lines are
saturated even when not black. We have also shown that in many cases
the shapes of the troughs are almost entirely due to changes in the
line of sight covering as a function of velocity, rather than to
differences in optical depth (Arav et~al.\ 1999b; de~Kool et~al.\
2001; Arav et~al.\ 2001b). Gabel et al.\ (2003) show the same effect in
the outflow troughs of NGC~3783. As a consequence, the column
densities inferred from the depths of the troughs are only lower
limits.

\clearpage

\vspace{-0.2cm}
\begin{figure} 
\plotfiddle{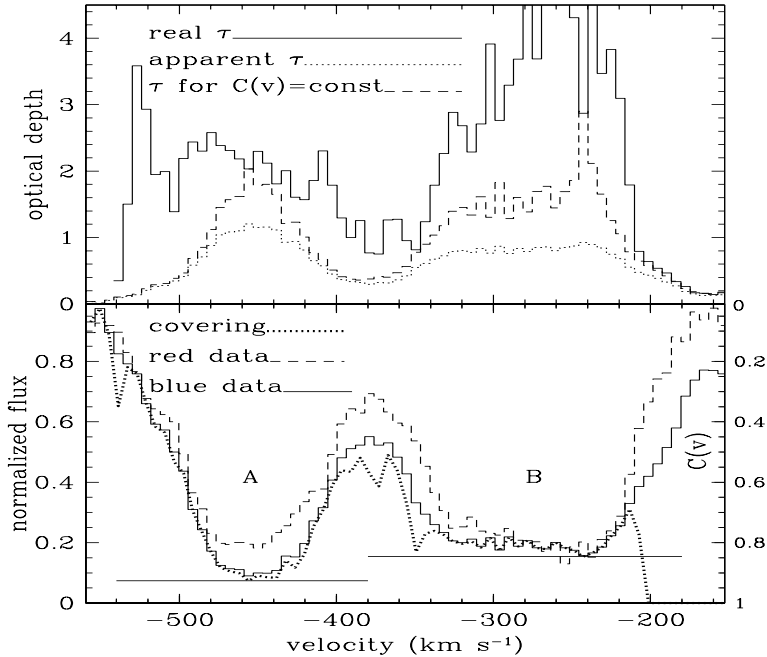}{3.1in}{0}{50}{35}{-150}{-20}
\caption{In the bottom panel we show the $C(v)$ solution  
for the \ovi\ troughs in Mrk 279. The straight lines below
the trough are our attempt to attach a single covering factor to each
of the troughs (covering fraction values are read from the right axis).
It is clear that the shape of trough A is dominated by $C(v)$.
The top panel shows three solutions for the optical depth:
Apparent $\tau$, which is derived from the residual intensity of the blue
doublet component assuming full coverage (i.e., the ISM and IGM case);
a solution (dashed line) assuming a constant covering (shown at the bottom
of troughs A and B); and the real $\tau$ using 
the $C(v)$ curve shown on the bottom panel.
}
\vspace{-0.2cm}
\end{figure}
\section{The importance of $C(v)$: \ovi\ troughs in Mrk 279}
\vspace{-0.1cm}
Mrk 279 was observed by FUSE for a combined 92 ksec on December 1999
and January 2000.  The source was in a high state, rivaling the
highest UV flux state of any Seyfert outflow target.  Several factors
combined to make these data arguably the highest quality FUSE
observation of an AGN outflow: Non-blending of the \ovi\ trough, the
length of exposure at a very high flux level, very low Galactic column
along the sight-line, and insignificant narrow emission line.  In
figure 1 we show the \ovi\ doublet data, covering fraction solution
and three different extractions of the optical depth. We
find that the shape of trough A is almost exclusively determined by
$C(v)$.  This feature manifests itself in the $\tau(v)$ solution.  The
column density extracted from the real solution is {\bf four times
larger} than the one we obtain from a simple inversion of the blue
doublet-component data.  Moreover the real column density is {\bf
three times larger} than the one we extract using a constant covering
fraction for this outflow component.  We also observe that the shape
of $\tau(v)$ is very different if we assume a constant covering
fraction instead of the actual $\tau(v)$.  Underestimating ionic
column densities by a factor of a few can translate to a much larger
discrepancy in the solution for the ionization parameter ($U$) and
total Hydrogen column density ($N_H$) of the outflow component.  As
shown in Arav (2001b) a five fold change in inferred ionic column
density can result in a 10 times higher $U$ and a 100 times higher
$N_H$.

\clearpage

%\begin{figure}    
%\plotone{cloudvsoutflow2.eps}
%\caption{My EPS graphic.}
%\end{figure}

\vspace{-0.2cm}
\section{Clouds vs. Dynamical Disk Outflows}
\vspace{-0.2cm}

\begin{figure}  
\plotfiddle{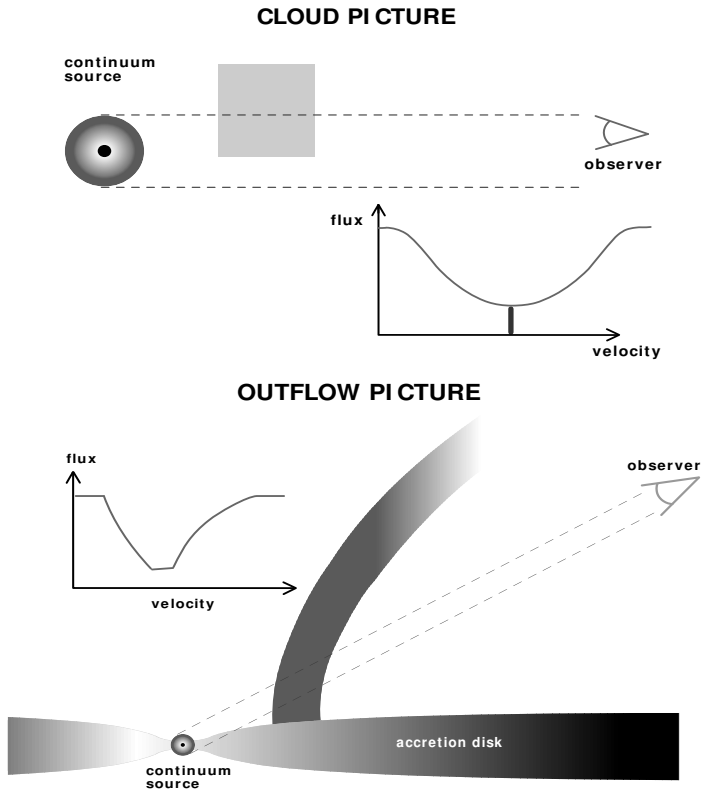}{3.9in}{0}{50}{45}{-130}{00}
\caption{ Possible geometries for the distribution of absorbing gas in
AGN outflows: Top, the traditional picture of a ``cloud,'' which only
covers part of the source.  Bottom, a dynamical outflow picture, which
takes acceleration and kinematic effects into consideration.  } 
\end{figure}

The
``cloud'' picture of absorption features associated with AGN outflows
was developed from the traditional ISM absorber model: A concentration
of gas with thermal velocity distribution.  Since the width of the
outflow absorption troughs were found to be 10--1000 times the expected
thermal width, a ``turbulent'' velocity broadening was invoked.  We
point out that this ``turbulent'' broadening is problematic since it
suggests internal motions with large Mach numbers which should result
in shocks and the destruction of the absorption ionic species.  In
addition, once it was realized that many of the troughs are saturated
but not black, a partial geometrical covering of the emission source
was invoked, as illustrated in the top part of figure 2.  The main
problem with this picture is its inability to explain the velocity
dependence of the covering fraction.  A cloud sitting in front of an
emission source should have a velocity independent covering fraction.
Furthermore, the cloud picture is hard pressed to explain the
observed detached troughs where the absorption starts from non zero
velocity.

We advocate the dynamical outflow scenario shown on the bottom of 
figure 2.  As we  show below, this model naturally
explains the strong velocity dependence of the covering fraction, the 
detached trough phenomenon and the highly super-thermal width of the 
trough. All these attributes are connected to
invoking the simplest reservoir for the outflowing material:
The accretion disk around the black hole. This scenario fits well with the
global AGN atmosphere model of Elvis (2000).

\clearpage

\begin{figure}  
\plotfiddle{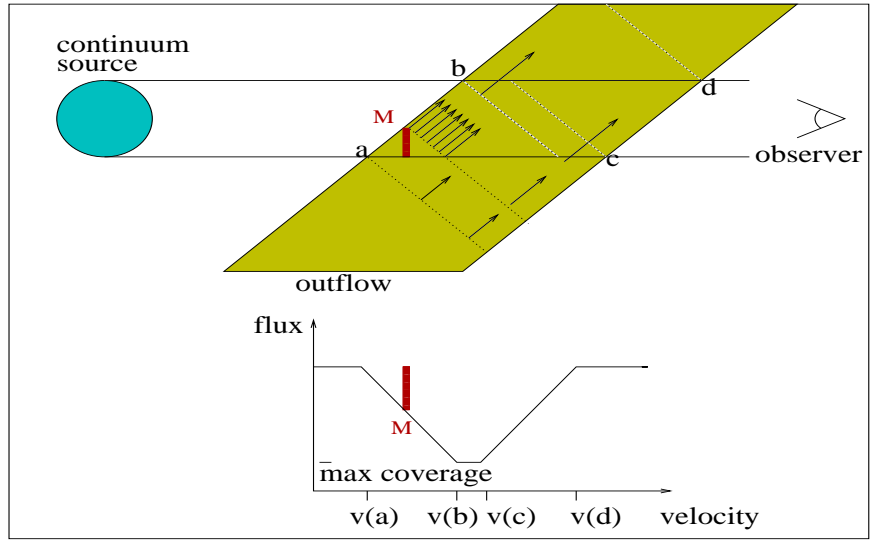}{2.8in}{-90}{50}{40}{-200}{230}
\caption{Dynamical formation model for a 
covering fraction dominated absorption trough }  
\label{in_out}
\vspace{-0.5cm}
\end{figure}

In figure 3 we illustrate how an accelerating outflow can
produce an absorption trough whose shape is totally determined by
$C(v)$.  We make two plausible assumptions: 1) The flow is at a small
angle to the line of sight (i.e. the flow is not purely radial).  2)
The flow is accelerating as it crosses the cylinder of sight, which
contains all the lines of sights from the continuum source to the
observer. For simplicity we also assume a 2-D geometry where the
velocity is constant perpendicular to the velocity vector.  A covering
fraction dominated absorption trough is produced for a fully opaque
flow in a given resonance line (e.g., \Lya).  Prior to the flow
reaching point {\bf a} it does not intersect the cylinder of sight,
therefore we do not see any absorption at $v<v_a$.  This scenario
gives a natural explanation to the widely observed phenomenon of
detached troughs. As the flow continues to accelerate, it gradually
covers a larger fraction of the cylinder of sight.  For example, at
point {\bf M} the flow covers roughly one third of the cylinder of
sight with material moving at $\vy{v}{M}$.  Even if the resonance line
opacity is very large at $\vy{v}{M}$ the absorption trough only dips
to $1/3$ below the continuum flux. $C(v)$ gradually increases until
$v_b$, at which point a maximum coverage has been reached. The maximum
coverage can be smaller than one in cases where there is an additional
extended emission source (e.g., the broad emission line region). At
$v_c$ the flow reaches the last point of maximum coverage and
thereafter the coverage decreases gradually. {\bf A fully opaque outflow
produces an absorption trough purely from velocity dependent covering
factor.}

\vspace{0.2cm}

\centerline {\bf References}

\medskip
\footnotesize

\hspace*{-0.44in}
\begin{tabular}{ll}

Arav, N., 1997, ASP 128, p. 208         & Elvis, M., 2000, ApJ, 545, 63 \\
Arav, N., et al., 1999a, ApJ, 516, 27   & Gabel, J. R., et al.,2003, ApJ, 595, 120 \\
Arav, N., et al., 1999b, ApJ, 524, 566  & Ganguly, R., et al., 1999, AJ, 117, 2594 \\
Arav, N., et al., 2001a, ApJ, 546, 140  & Kaastra, J. S., et al., 2000, A\&A, 354L, 83 \\
Arav, N., et~al., 2001b,  ApJ, 561, 118 & Kaspi, S., et al., 2000, ApJ, 535L, 17 \\
Barlow, T. A., 1997  ASP 128, p. 13     & Korista, T. K., et al., 1993, ApJS, 88, 357 \\
Churchill, C. W., et al., 1999, AJ, 117, 2573 & Kriss, G. A., et al., 2000, ApJL, 538, 17 \\
Crenshaw, D. M., et al., 1999 ApJ, 516, 750  & Telfer, R.C.,  et al., 1998, ApJ,  509, 132 \\
de~Kool, M., et al., 2001, ApJ, 548, 609 & Weymann, R. J.,  et al., 1991, ApJ, 373, 23  \\

\end{tabular}

\end{document}